\documentclass[12pt]{article}\usepackage[dvips]{graphicx}
\usepackage{latexsym}
\usepackage{amssymb}

\newcommand{\C}{\mathbb{C}}
\newcommand{\CP}{\mathbb{CP}}

\newcommand{\R}{\mathbb{R}}

\newcommand{\Z}{\mathbb{Z}}

\renewcommand{\d}{\mathrm{d}}

\newcommand{\koniec}{\begin{flushright}  $\Box $ \end{flushright}}
\def\be{\begin{equation}}
\def\ee{\end{equation}}

\def\u{\bf u}

\def\u{\bf u}

\def\Om{\Omega}

\def\p{\partial}
\def\ov{\overline}
\def\k{\kappa}

\def\a{\alpha}
\def\ll{\lambda}

\newtheorem{theo}{Theorem}[section]
\newtheorem{prop}[theo]{Proposition}

\begin{document}
\title{\vskip -70pt
\begin{flushright}
{\normalsize DAMTP-2006-39} \\
\end{flushright}
\vskip 80pt
{\bf Topology and Energy of Time Dependent Unitons}\vskip 20pt}
\author{Maciej Dunajski\thanks{email M.Dunajski@damtp.cam.ac.uk}\\[15pt]
and\\[10pt]
Prim Plansangkate\thanks{email P.Plansangkate@damtp.cam.ac.uk}
\\[15pt]
{\sl Department of Applied Mathematics and Theoretical Physics} \\[5pt]
{\sl University of Cambridge} \\[5pt]
{\sl Wilberforce Road, Cambridge CB3 0WA, UK}
}
\date{}
\maketitle
\begin{abstract}
{We consider a class of time dependent finite energy multi--soliton solutions
of the $U(N)$ integrable chiral model in $(2+1)$ dimensions.
The corresponding extended solutions of the associated linear problem
have a pole with arbitrary
multiplicity in the complex plane of the spectral parameter. Restrictions of
these extended solutions to any spacelike plane in $\R^{2,1}$ have trivial
monodromy and  give rise to  maps from a three sphere to $U(N)$.
We demonstrate that the total energy of each multi--soliton is
quantised at the classical level and given by the third homotopy
class of the extended solution.  This is the first
example of a topological mechanism explaining classical energy
quantisation of moving solitons.}
\end{abstract}
\newpage

\section{Introduction}
\setcounter{equation}{0}

The fact that the allowed energy levels of some physical systems can take only
discrete values has been well known since the the early days of
quantum theory. The  hydrogen atom and the harmonic oscillator are two well
known examples. In these two cases
the boundary conditions imposed on the
wave function imply discrete spectra of the Hamiltonians.
The reasons are therefore global.

The quantisation of energy can also occur at the classical level in
nonlinear field theories if the
energy of a smooth field configuration is finite.
The reasons are again global, but one needs more subtle ideas from topology
to understand what is going on.
The potential energy of static  soliton solutions
in the Bogomolny limit of certain field theories
must be proportional to integer homotopy classes of smooth maps.
The details depend on the model: In pure gauge theories the energy of solitons
satisfying the Bogomolny equations is given by one of the Chern numbers of the
curvature. In scalar 2+1 dimensional sigma models, allowed
energies of Bogomolny solitons
are given by elements of $\pi_2(\Sigma)$, where the manifold $\Sigma$
is the target space. In both cases the
boundary condition are used to show that the finite energy configurations
extend to the compactification of space. See \cite{MS04}
for a detailed exposition of these constructions.

The situation is different for moving solitons: The total energy is
the sum of kinetic and potential terms, and the Bogomolny bound
is not saturated. One expects that the moving (non-periodic)
solitons will have continuous
energy. Attempts to construct theories with quantised total energy
based on compactifying the time direction are physically unacceptable,
as they lead to paradoxes related to the
existence of closed time--like curves.
A soliton moving along such curve could eventually reach its own past
thus opening possibilities to
sinister scenarios usually involving a death of somebody's great grandparents.

In a recent publication \cite{IM04} Ioannidou and Manton made the
surprising observation  that the total  energy of the
time--dependent $SU(2)$ two--uniton solution of Ward's $2+1$
dimensional chiral model \cite{W88,W95} is quantised in the units
of $8\pi$ when the pole of the corresponding extended solution is
at $ \pm i$. They have shown that the two--uniton energy density
calculated at any instant of time $t$ is the same as the energy
density of a static $\CP^3$ multi--lump with a parameter $t$. The
total (potential) energy of the latter model is  quantised
\cite{Zak} which leads to the total (kinetic+potential) energy
quantisation of the time--dependent unitons. The quantisation was
also obtained by  Lechtenfeld and  Popov \cite{LP01a,LP01b} whose
method was based on large time asymptotic analysis.

One expects that there are deeper topological reasons for this
quantisation, and the purpose of this paper is to show that this
is indeed the case.

The Ward chiral model is
\be \label{Wardeq}
(J^{-1}J_t)_t-(J^{-1}J_x)_x-(J^{-1}J_y)_y-[J^{-1}J_t,
J^{-1}J_y]=0, \ee
where $J:\R^{3}\longrightarrow U(N)$, and
$x^{\mu}=(t, x, y)$ are coordinates on $\R^3$ such that the line
element is $\eta=-\d t^2+\d x^2+\d y^2$. Here we use notation $
J_\mu := \p_\mu J $. The equations are not fully Lorentz
invariant, as the commutator term fixes a space--like direction. A
positive definite conserved energy functional for (\ref{Wardeq})
is
\be \label{energy} E=\int_{\R^2}{\cal E}\d x\d y, \ee where the
energy density  is given by
\be \label{density} {\cal
E}=-\frac{1}{2}\mbox{Tr}((J^{-1}J_t)^2+(J^{-1}J_x)^2+(J^{-1}J_y)^2).
\ee

The integrability of  (\ref{Wardeq})
allows a  construction of explicit static and also time--dependent
solutions by twistor or inverse--scattering methods \cite{W88,W90}.
There are time--dependent solutions
with non--scattering solitons \cite{W88}, and also solitons that
scatter \cite{W95}. A class of scattering solutions to (\ref{Wardeq}) is
given by so called time--dependent unitons
\be
\label{n_uniton}
J(x, y, t)=M_1 M_2\ldots M_n,
\ee
where the unitary matrices $M_k, k=1, \ldots, n$ are given by
\be\label{M_k}
M_k=i\Big({\bf 1}-
\Big( 1- \frac{\mu}{\bar \mu} \Big) R_k  \Big), \qquad  R_k \equiv \frac{q_{k}^*\otimes q_{k}}{||q_{k}||^2}.
\ee Here $\mu\in\C \backslash \R$ is a non-real constant and
$q_k=(1, f_{k1}, ... , f_{k(N-1)}) \in\C^N$, with $k=1, \ldots,
n$, are vectors whose components $f_{kj}=f_{kj}(x^{\mu})\in \C$
are smooth functions which tend to a constant at spatial
infinity\footnote{ The matrix $R_k$ is a hermitian projection
satisfying $(R_k)^2=R_k$, and the corresponding $M_k$ is a
Grassmanian embedding of $\CP^{N-1}$ into $U(N)$. The results in
this paper apply to the more general class of unitons obtained
from the complex Grassmanian embeddings of $Gr(K, N)$ into the
unitary group. For $\mu$ pure imaginary, a complex $K$--plane
$V\subset \C^N$ corresponds to a unitary transformation
$i(\pi_V-\pi_{V^{\perp}})$, where $\pi_V$ denotes the hermitian
orthogonal projection onto $V$. The formula (\ref{M_k}) with $\mu=i$
corresponds to $K=1$ where $Gr(1, N)=\CP^{N-1}$.}.

If $n=1$ then $q_1$
is holomorphic and rational in  $\omega= x +   \frac{1}{2} \mu (t+y)
+ \frac{1}{2} \mu ^ {-1}  (t-y) $  \cite{W88}. Note that if $\mu = \pm
i$, $q_1$
does not depend on $t$, and the  corresponding $1$--uniton
is static. If $n>1$ $q_1$ is still holomorphic and rational in $\omega$, but
$q_2, q_3, ... $ are not holomorphic. The exact form
of this functions is known explicitly for $n=2, 3$ \cite{W95, I96} for the case $N=2$.
For $n>3$ the B\"acklund transformations \cite{IZ98,DT04}
can be used to determine the $f$s recursively.
The total energy (\ref{energy}) of $n$--uniton  solutions is finite.

In general the finiteness of $E$
is ensured  by imposing the boundary condition (valid for all $t$)
\be
\label{assympt}
J=J_0+J_1(\varphi)r^{-1}+O(r^{-2})\qquad
\mbox{as}\qquad r\longrightarrow \infty,\qquad x+iy=re^{i\varphi}
\ee
and so for a fixed value of $t$ the matrix $J$ extends to a
map from $S^2$ (conformal
compactification of $\R^2$) to $U(N)$. The homotopy group
$\pi_2(U(N))=0$, so there is no topological
information in $J$ defined on $\R\times S^2$ which could be related to the
total energy.
We shall nevertheless show that the energy of (\ref{n_uniton}) is
quantised, and given by the third homotopy class
of the extended solution to (\ref{Wardeq}). The existence of this
extended
solution is linked to the complete integrability of (\ref{Wardeq})
and the associated Lax equations with the spectral parameter. The extended
solution also depends on this parameter, and hence is defined on $\R
^ 3 \times \CP ^1$.    Restricting it to a space--like plane
in $\R^3$ and an equator in a Riemann sphere of the spectral parameter
gives a map $\psi$, whose domain is $\R^2\times S^1$.  If $J$ is an
$n$-uniton solution (\ref{n_uniton}), the corresponding extended
solution satisfies stronger  boundary conditions which promote
$\psi$ to a map $S^3\longrightarrow U(N)$.
In the next Section we shall introduce the extended solution, and
impose boundary conditions on $J$ which are stronger than (\ref{assympt})
and in fact provide a coordinate--free characterisation of the uniton
solutions (\ref{n_uniton}).
In Section \ref{Section_3} we shall establish the following result:
\begin{theo}
\label{main_th}
The total energy of the $n$-uniton solution (\ref{n_uniton}) with complex number $\mu  = m e^{i \phi} $  is
quantised and equal to
\be
\label{conj1}
E_{(n)} = 4 \pi \Big( \frac{1+m^2}{m} \Big) |\sin ( \phi )| \; [\psi] ,
\ee
where for any fixed value of $t$
the map  $\psi:S^3 \longrightarrow U(N) $ is given by
\be
\psi = \prod _{k=n}^1 \Big({\bf 1}+\frac{\ov{\mu}-\mu}{\mu + \cot
\Big(\frac{\theta}{2} \Big)} R_k  \Big), \qquad \theta \in [0, 2 \pi],
\ee
and
\be
[\psi]=\frac{1}{24\pi^2}\int_{S^3}\mbox{Tr}\;((\psi^{-1}\d \psi)^3)
\ee
is an integer taking
values in $\pi_3(U(N))=\Z$.
\end{theo}

The  model (\ref{Wardeq}) is $SO(1, 1)$ invariant,
and in Section \ref{Section_3} it will be shown that the Lorentz
boosts correspond to rescaling $\mu$ by a real number. The rest frame
corresponds to $|\mu|=1$,  when the $y$-component of
the momentum vanishes. The $SO(1, 1)$ invariant generalisation of (\ref{conj1})
will be given by Theorem \ref{theo_inv}.
Energies of  soliton solutions more general than (\ref{n_uniton})
are briefly discussed in Section \ref{outlook}.

\section{Extended solution and its homotopy}

\subsection{Lax pair and trivial scattering}

The proof of Theorem (\ref{main_th}) relies on integrability of
(\ref{Wardeq}) and its Lax formulation, which we set up next.
Let $A=A_\mu\d x^{\mu}$ and $\Phi$ be a one--form and a function
on $\R^{2,1}$ with values in a Lie algebra of $U(N)$ determined up
to gauge transformations
\[
A\longrightarrow bAb^{-1} -\d b \; b^{-1}, \qquad
\Phi\longrightarrow b\Phi b^{-1}, \qquad b=b(x^\mu)\in U(N).
\]
The system of first order equations
\[
D\Phi=*F,
\]
where $D\Phi=\d\Phi+[A, \Phi]$ and $F=\d A+A\wedge A$,
gives the
integrability conditions $[L_0, L_1]=0$ for an overdetermined system of linear
equations
\be
\label{laxpair}
L_0\Psi:=(D_y+D_t-\ll(D_x+\Phi))\Psi=0,\qquad
L_1\Psi:=(D_x-\Phi-\ll(D_t-D_y))\Psi=0,
\ee
where $\Psi$ is an $GL(N, \C)$-valued function of $x^{\mu}$ and a complex
parameter $\ll\in \CP^1$, which satisfies the unitary reality condition
\[
\Psi(x^{\mu}, \ov{\ll})^*\Psi(x^{\mu}, {\ll})={\bf 1}.
\]
The matrix $\Psi$ is also subject to gauge transformation
$\Psi \longrightarrow b\Psi$.
The integrability conditions for (\ref{laxpair}) imply the
existence of a gauge $A_t=A_y$, and  $A_x=-\Phi$, and a matrix
$J:\R^{3}\longrightarrow U(N)$ such that
\[
A_t=A_y=\frac{1}{2}J^{-1}(J_t+J_y), \qquad
A_x=-\Phi=\frac{1}{2}J^{-1}J_x,
\]
and equations (\ref{Wardeq}) hold. Given a solution $\Psi$ to the
linear system (\ref{laxpair}) one can construct a solution to
(\ref{Wardeq}) by \be \label{JfromP} J(x^{\mu})=\Psi^{-1}(x^{\mu},
\ll=0) \ee and all solutions to (\ref{Wardeq}) arise from some
$\Psi$'s.  The detailed exposition of this is presented, for
example, in \cite{HSW99}.

Let us restrict $\Psi$ from $\R^{2,1} \times \CP^1$ to the
space-like plane $t=0$.  We shall also restrict the spectral
parameter to lie in the real equator $S^1 \subset \CP^1$
parameterised by $\theta$:
\be 
\label{restrict} \Psi
(t,x,y, \ll) \longrightarrow \psi(x, y, \theta) := \Psi (x, y, 0, -\cot
\frac{\theta}{2}), 
\ee
where
now $\psi: \R^2 \times S^1 \longrightarrow U(N)$ and we made
change of variable for real $\ll = -\cot{ ( \frac{\theta}{2}) }$.
Note that $\psi$ automatically satisfies 
\be
\label{holonomy2} (u^\mu D_\mu-\Phi)\psi=0, 
\ee 
where the operator anihilating $\psi$
is the spatial part of the Lax pair (\ref{laxpair}), given by
\[
\label{holonomy} \frac{\ll L_0+ L_1}{1+\ll^2}=u^\mu D_\mu-\Phi,
\qquad \mbox{where}\qquad {\u}=\Big(0, \frac{1-\ll^2}{1+\ll^2},
\frac{2\ll}{1+\ll^2}\Big) = (0, -\cos{\theta}, -\sin{\theta}).
\]

We impose the `trivial scattering' boundary
condition  \cite{An,W98} 
\be \label{trivial_scatt} \psi (x,y, \theta)
\longrightarrow \psi_0(\theta) \qquad \mbox{as} \qquad r
\longrightarrow \infty, 
\ee 
where $\psi_0(\theta)$ is an
$U(N)$-valued function on $S^1$. We shall now demonstrate that this enables
us to extend $\psi$ to a map from $S^3$ to $U(N)$.

First note that (\ref{trivial_scatt}) implies the existence of the
limit of $\psi$ at spatial infinity for all values of $\theta$,
while the finite energy boundary condition (\ref{assympt}) only
implies the limit at $\theta = \pi$.  Thus the condition
(\ref{trivial_scatt}) extends the domain of $\psi$ to $S^2\times
S^1$. However, it turns out that (\ref{trivial_scatt}) is also a
sufficient condition for $\psi$ to extend to the suspension $SS^2
= S^3$ of $S^2$.  This can be seen as follows. The domain $S^2
\times S^1$ can be considered as $S^2 \times [0,1]$ with  $\{0 \}$
and $\{1 \}$ identified. Recall that a suspension $S X$ of a
manifold $X$ is the quotient space \cite{topol}
\[
S X=([0, 1]\times X)/((\{0\}\times X)\cup (\{1\}\times X)),
\]
This definition is compatible with spheres in the sense that $SS^d
= S^{d+1}$.

Now the only condition $\psi$ needs to fulfils for the suspension
is an equivalence relation between all the points in $S^2 \times
\{ 0 \}$, since such relation for  $S^2 \times \{ 1 \}$ will
follow from the identification of $\{0\}$ and $\{1\}$. 
This  equivalence can be achieved by choosing  a gauge
 \be
\label{gauge1} \psi(x, y, 0)={\bf 1}. \ee
Therefore $\psi$
extends to a map from $SS^2 = S^3$ to $U(N)$ if it satisfies the
zero scattering boundary condition.

\begin{figure}
\caption{ Suspension and Reduced Suspension}
\label{suspension_fig}
\begin{center}
\includegraphics[width=10cm,height=5cm,angle=0]{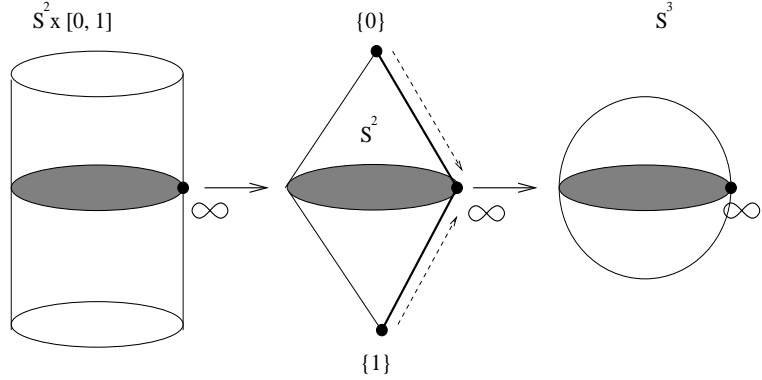}
\end{center}
\end{figure}

In addition, after fixing the gauge (\ref{gauge1}), there is still
some residual freedom in $\psi$ given by
\be
\label{additional_f}
\psi \longrightarrow   \psi K,
\ee
where $K=K(\theta, x, y)\in U(N)$ is anihilated by $u^{\mu}\p_{\mu}$.
Setting  $K = ({\psi_0}(\theta))^{-1} $ results in 
\be
\label{gauge2} \psi(\{\infty\}, \theta)={\bf 1}. 
\ee 
The gauge (\ref{gauge2}) picks a base point $\{x_0= \infty \} \in S^2$, and
this implies that the zero scattering condition is also sufficient
 for $\psi$ to extend to the reduced suspension of $S^2$, given
by
\[
S_{red}S^2 =([0, 1]\times S^2)/((\{0\}\times S^2)\cup (\{1\}\times
S^2) \cup ([0,1] \times \{ x_0 \})).
\]
This is also homeomorphic to $S^3$.  The idea of (reduced)
suspension is illustrated in (Fig. \ref{suspension_fig}).

Now let us justify the term `trivial scattering' in
(\ref{trivial_scatt}).  Consider equation (\ref{holonomy2}) and
restrict it to a line $ (x,y) = (x_0-\sigma \cos{\theta},
y_0-\sigma \sin{\theta}), \sigma\in\R$.  Now (\ref{holonomy2})
becomes an ODE describing the propagation of
\[
\psi=\psi(x_0-\sigma \cos{\theta}, y_0-\sigma \sin{\theta}, \theta)
\]
along the oriented line through $(x_0, y_0)$ in $\R^2$.  
We can choose a gauge such
that
\[
\lim_{\sigma\rightarrow -\infty}\psi={\bf 1},
\]
and define the scattering matrix $S:TS^1\rightarrow U(N)$ on the
space of oriented  lines in $\R^2$ as
\be \label{scattering}
S=\lim_{\sigma\rightarrow \infty}\psi. \ee
The trivial scattering
condition (\ref{trivial_scatt}) then implies this matrix is
trivial,
\be 
\label{trivscatt} S={\bf 1}.
\ee

As we have explained, the boundary condition (\ref{assympt}) and
(\ref{trivial_scatt}) imply that for each value of $\theta$ the
function $\psi$ extends to a one-point compactification $S^2$ of
$\R^2$.  The straight lines on the plane are then replaced by the
great circles, and in this context the trivial scattering
condition implies that the differential operator
$u^{\mu}D_{\mu}-\Phi$
has trivial monodromy along the compactification $S^1=\R\cup\{\infty\}$ of
a straight line parametrised by $\sigma$.

\subsection{Topology of extended solution}
In the last subsection we have explained that
we can regard $\psi$ as a map from $S^3$ to $U(N)$.  All such
maps are characterised by their homotopy type \cite{topol} \be
\label{degree1}
[\psi]=\frac{1}{24\pi^2}\int_{S^3}\mbox{Tr}((\psi^{-1}\d \psi)^3).
\ee The element $[\psi]$  is an integer taking values in
$\pi_3(U(N))=\Z$, and is invariant under continuous deformations
of $\psi$.

In the next section we will need the following result: Let
$g_1$ and $g_2$ be maps from $S^3$ to  $U(N)$ and let
$g_1g_2:S^3\longrightarrow U(N)$ be given by
\[
g_1g_2(x):=g_1(x)g_2(x),  \qquad x\in S^3,
\]
where the product on the RHS is the point-wise group multiplication. Then
\be
\label{sum_of_degrees}
[g_1g_2]=[g_1]+[g_2].
\ee
This is because
\[
\mbox{Tr}[(({g_1g_2})^{-1}\d (g_1g_2))^3]
=\mbox{Tr}[({g_1}^{-1}\d{g_1})^3+(g_2^{-1}\d g_2)^3]
+\d \beta,
\]
where $\Om$ is a two--form, and so $\d \beta $ integrates to $0$ by
Stokes' theorem. This was explicitly demonstrated by Skyrme
\cite{S62} in the case of $SU(2)$.

Rather than exhibiting the exact form of $\beta$ we shall use the
following general argument.
The higher homotopy groups $\pi_d(G)$ of a Lie group $G$
are abelian, and the group multiplication in $G$ induces the addition
in the homotopy groups: if
$g_1$ and $g_2$ are maps from $S^d$ to  $G$
then the homotopy class of the map $g_1g_2:S^d\longrightarrow G$ defined by
the group multiplication is the sum of homotopy classes of $g_1$ and $g_2$.
The proof of this
is presented for example in \cite{topol} and essentially follows
the proof that the fundamental group of a topological group is abelian.
Now $\pi_3(G)=\Z$ for any compact simple Lie group. If $G=SU(2)$ this
result just reproduces the calculation done by Skyrme
as  two continuous maps from $S^3$ to itself are homotopic
iff they have the same topological degree. Theorem \ref{main_th}
holds for unitons with value in $G=U(N)$, where $[\psi]$ in
(\ref{conj1}) is the sum  of homotopy classes which arise from
integrals of elements of $H^3(G)$. To find out a homotopy class of a map
$\psi$ we can use the formula (\ref{degree1}), where the integrand
is a left--invariant three--form on the group manifold pulled back to $S^3$.
This is because
$\pi_3(G)$ is isomorphic to the integral homology group
$H_3(G, \Z)$, and the RHS of (\ref{degree1}) coincides with the homology class
of  the cycle $\psi(S^3)\subset G$.

We remark that some
part of this topological data is encoded in the Ward equation
(\ref{Wardeq}), which can be regarded as an ordinary chiral model
with torsion \cite{W88T}.
Any compact semi--simple
group $G$ admits a connection which parallel propagates
left--invariant vector fields. This connection is flat, but necessarily
has torsion $T$. The torsion is totally anti--symmetric, thus
giving a preferred  three--form in the  third cohomology group
which can then be pulled back to $S^3$.
The first order commutator term in (\ref{Wardeq}) can be
rewritten as
\[
\epsilon^{\mu\nu}[J^{-1}\p_\mu J, J^{-1}\p_{\nu} J],
\]
where $\epsilon^{\mu\nu}$ is a totally antisymmetric constant matrix. In our
case
$\epsilon^{\mu\nu}=\varepsilon^{\mu\nu\a}V_\a$, where $V=(0, 1, 0)$
is the space--like unit vector and  (\ref{Wardeq}) takes the form
\[
(\eta^{\mu\nu}+\epsilon^{\mu\nu})(\p_{\mu}(J^{-1}\p_{\nu} J))=0.
\]
The choice of $V$ reduces the symmetry group
of (\ref{Wardeq}) down to $SO(1, 1)$. The momenta $P_t=E$, and $P_y$ are
well defined, and conserved for (\ref{Wardeq}).

The commutator term can be obtained from a Lagrangian density
$\epsilon^{\mu\nu}(\p_\mu \xi^i)(\p_\nu \xi^j)e_{ij}(\xi)$,
where the two--form $e$ is a local potential for the torsion
$\d e=T$, and $\xi^j$ are local coordinates on $G$. The  two--form $e$
is defined only locally in $G$.

\section{Time dependent unitons and energy quantisation}
\label{Section_3} A class of  extended solutions which satisfy the
trivial scattering condition (\ref{trivial_scatt}) give rise to
the $n$-uniton solutions defined in (\ref{n_uniton}).  These
extended solutions factorise into the so called $n$-uniton factors
\cite{W95} 
\be \label{n-uniton} \Psi=G_nG_{n-1}\; \ldots\;
G_1,\quad \mbox{where}\quad G_k=\Big({\bf
1}-\frac{\ov{\mu}-\mu}{\ll-\mu} R_k \Big)\in GL(N, \C), \quad R_k
= \frac{q_{k}^*\otimes
  q_{k}}{||q_{k}||^2}.
\ee
Here $q_k=q_k(x, y, t)\in\C^N,  k=1, \ldots, n$,
and $\mu$ is a non-real constant.
The terminology here is rather confusing,
as the maxima of the energy density of the
corresponding soliton solutions of (\ref{Wardeq}) do physically scatter.
The exact form of $q_k$s is determined
from (\ref{laxpair}) by demanding that the expressions
\be
\label{at_most_lin}
(\p_x\Psi-\ll(\p_t-\p_y)\Psi)\Psi^{-1}, \qquad\mbox{and}\qquad
((\p_t+\p_y)\Psi-\ll\p_x\Psi)\Psi^{-1}
\ee
are independent of $\ll$. In practice one determines the $q_k$s
by a limiting procedure  from solutions of a
Riemann problem with simple poles \cite{W88}.

The restricted map $\psi$ (\ref{restrict}) corresponding to
(\ref{n-uniton}) is given by
\be \label{simple_elements}
\psi=g_ng_{n-1}\; \ldots\; g_1, \qquad \mbox{where} \qquad g_k =
{\bf 1}+ \frac{\bar \mu-\mu}{\mu + \cot \Big( \frac{\theta}{2}
\Big)}R_k \in U(N), \ee
where $ \ll = -\cot \Big( \frac{\theta}{2}
\Big) \in S^1 \subset \CP ^1$ as before and all the maps are
restricted to
 the $t=0$ plane.  Each element $g_k$ has the limit at spatial
 infinity for all values of $\theta$.
\[
g_k (x,y, \theta) \longrightarrow g_{0k}(\theta) = {\bf 1}
+ \frac{\ov{\mu}-\mu}{\mu + \cot \Big( \frac{\theta}{2} \Big)}R_{0k}
\qquad \mbox{as} \qquad x^2 + y^2 \longrightarrow \infty.
\]
The existence of the limit at spatial infinity 
$R_{0k}=\lim_{r\rightarrow\infty}R_k(x,y)=const$ 
is guaranteed by the finite energy
condition (\ref{assympt}). Hence $\psi$ (\ref{simple_elements})
satisfies the trivial scattering condition (\ref{trivial_scatt})
and extends to a map from $S^3$ to $U(N)$.  The scattering 
matrix\footnote{ Novikov \cite{Nov02} has
demonstrated that given a scattering matrix on the space of
oriented lines in $\R^D$ with $D>2$ it  is always possible to
reconstruct the gauge potential and the Higgs field on $\R^D$ by
means of a non--Abelian  inverse Radon transform. The non--trivial
initial data for the time dependent $n$--unitons (\ref{n-uniton})
has trivial scattering matrix which shows that the inversion is
not in general possible if $D=2$.} (\ref{scattering}) is $S = {\bf 1}$.

 Note, however, that the $g_k$s
and $\psi$ in (\ref{simple_elements}) only extend to the ordinary
suspension of $S^2$.  One needs to perform the transformation
(\ref{additional_f}) with $K= \prod _{k=1}^n g_{0k}^{-1} $ for $\psi$ to
extend to the reduced suspension of $S^2$.  We shall use $\psi$ as
in (\ref{simple_elements}), because 
(\ref{sum_of_degrees}) and 
$\pi_1(U(N))=0$ imply that
the transformation
(\ref{additional_f}) does not contribute to the degree and $[K(\theta)
\psi] = [\psi]$.

\begin{prop}
\label{homotopy_prop} The third homotopy class of $\psi$ is given
by \be \label{homotopy_of_psi} [\psi] =  \pm \frac{i}{2\pi}
\int_{\R^2} \sum^{n}_{k=1}  \mbox{Tr} (R_k[\p_x R_k , \p_y
  R_k]) \d x \d y \quad \left\{ \begin{array}{ll}
                       0 < \phi < \pi \\
                       \pi < \phi < 2 \pi,
                       \end{array}
                       \right.
\ee where $\mu = me^{i \phi}$.
\end{prop}
{\bf Proof.}
The recursive application of (\ref{sum_of_degrees}) implies that
\[
[\psi] = \sum^{n}_{k=1} [g_k].
\]
Using (\ref{degree1}), with $z=x+iy$,
\begin{eqnarray*}
\label{calculation1}
[g_k]&=&
\frac{1}{8\pi^2}\int_{S^1 \times\R^2 } \mbox{Tr}{(g_k^{-1}\p_{\theta}g_k\;
[g_k^{-1}\p_z g_k, g_k^{-1}\p_{\ov z}g_k]})\;\d \theta\wedge\d z\wedge
\d \ov{z} \\
&=&\frac{1}{16\pi^2} \Theta(\mu) \int_{\R^2} \mbox{Tr} (R_k
[\p_zR_k, \p_{\ov{z}}R_k])
\;\d z\wedge
\d \ov{z}\nonumber,
\end{eqnarray*}
where
\begin{eqnarray*}
\Theta(\mu) &=& \int_{0}^{2\pi} \frac{ (\bar \mu -\mu)^3 \sin^2 \Big(
  \frac{\theta}{2} \Big)} { \Big( |\mu|^2 + (1-|\mu|^2) \cos^2 \Big( \frac{\theta}{2} \Big) + (\mu+\bar
  \mu) \cos \Big(\frac{\theta}{2} \Big) \sin \Big(\frac{\theta}{2}
  \Big) \Big) ^2 }  \d \theta \\
&=& \pm 8\pi i \quad \left\{ \begin{array}{ll}
                       0 < \phi < \pi \\
                       \pi < \phi < 2 \pi.
                       \end{array}
                       \right.
\end{eqnarray*}
Hence, changing to the $(x,y)$
coordinates, we obtain
\be
\label{form_of_g}
[g_k] = \pm \frac{i}{2\pi} \int_{\R^2} \mbox{Tr} (R_k[\p_x R_k , \p_y
  R_k]) \d x \d y \quad \left\{ \begin{array}{ll}
                       0 < \phi < \pi \\
                       \pi < \phi < 2 \pi.
                       \end{array}
                       \right.
\ee
Therefore, the third homotopy class of $\psi$ is given by
(\ref{homotopy_of_psi}). \koniec
The proof of Theorem (\ref{main_th}) makes use of the above
Proposition and a recursive procedure of adding unitons to a given
solution which we shall now explain.
Let $\Psi$ be an extended solution to the Lax pair (\ref{laxpair})
corresponding to a solution $J$, which satisfies (\ref{Wardeq}).
Set
\be \label{hatpsi} \hat{\Psi}=G \Psi=\Big({\bf
1}-\frac{\ov{\mu}-\mu}{\ll-\mu}R  \Big) \Psi,\qquad
\hat{J}=\hat{\Psi}^{-1}|_{\ll=0}=JM, \ee
where $M$ is of the form
(\ref{M_k}), up to a constant phase factor which is irrelevant.
The matrix $\hat{\Psi}$ will be an extended solution if
expressions (\ref{at_most_lin}) with $\Psi$ replaced by
$\hat{\Psi}$ are independent of $\ll$. This leads to the
B{\"a}cklund relations \cite{IZ98, DT04}. These are first order
PDEs for $M$, which can be regarded as a generalisation of
Uhlenbeck's method of adding unitons for harmonic maps \cite{U89}.
In terms of the hermitian projection $R$, these PDEs are
\begin{eqnarray}
\label{backlund}
R(R_t-J^{-1}J_t({\bf 1}-R)) &=& B \\
RR_t &=& C, \nonumber
\end{eqnarray}
where
\begin{eqnarray}
B &=& (\mu R_x - R_y + RJ^{-1}J_y)({\bf 1}-R) \nonumber \\
C &=& \frac{1}{\mu}((\mu R_y + R_x - RJ^{-1}J_x)({\bf 1}-R)).
\nonumber
\end{eqnarray}
{\bf Proof of Theorem \ref{main_th} } We first consider a solution
of the form $ \hat J = JM $, where $J$ is an arbitrary solution of
(\ref{Wardeq}). Noting that $M$ is unitary and writing it in terms
of $R$, the difference between the energy densities
(\ref{density}) of $\hat{J}$ and $J$ is given by
\be  \label{DeltaE} \Delta \mathcal{E} \equiv \hat{\mathcal{E}} -
\mathcal{E} = \sum_{a}\mbox{Tr}\;( \k \bar \k R_aR_aR + \k ({\bf
1}-\bar \k R) J^{-1}J_aR_a ), \ee where $a$ stands for $(t,x,y)$,
$\hat \mathcal{E}$ and  $\mathcal{E}$ are the energy densities of
$\hat J$ and $J$ respectively
and $\k = \Big( 1 - \frac{\mu}{ \bar \mu } \Big) $.

Multiplying the relations (\ref{backlund}) and their hermitian
conjugates yields the following identities
\begin{eqnarray} \label{identity}
\mbox{Tr}(R_tR_tR) &=& \mbox{Tr}(CC^*) \nonumber\\
\mbox{Tr}(J^{-1}J_tR_t) &=& \mbox{Tr}(CB^* - BC^*)\nonumber \\
\mbox{Tr}(RJ^{-1}J_tR_t) &=& \mbox{Tr}( (C-B)C^* ).
\end{eqnarray}
The terms involving time derivatives in (\ref{DeltaE}) are of the
form $ R_tR_tR$, $J^{-1}J_tR_t$ and $RJ^{-1}J_tR_t$, which, by
(\ref{identity}), can be written in terms of the spatial
derivatives only.  Thus by direct substitution and some
rearrangements (\ref{DeltaE}) becomes
\[
\Delta \mathcal{E} = -\frac{\k}{\mu} \mbox{Tr} \Big( (1+|\mu|^2)
  R[R_x,R_y] + \cal{T} \Big),
\]
where $ {\mathcal{T}} = \p_x (RJ^{-1}J_y) - \p_y (RJ^{-1}J_x)$
gives no contribution to the difference in the energy functionals
of $\hat J$ and $J$.  This is because
 \begin{eqnarray*}
&&\mbox{Tr}\int_{\R^2} \mathcal{T} \d x\wedge \d y=
\lim_{r\rightarrow\infty}
\int_{D_r} \d(\mbox{Tr}(RJ^{-1}\d J))\\
&=& \lim_{r\rightarrow \infty}\oint_{C_r} \mbox{Tr}(RJ^{-1}\d J) =
\mbox{Tr}\Big(\lim_{r\rightarrow \infty}\oint_{C_r} (JR)^* \d J\Big)\\
&\leq& \lim_{r\rightarrow \infty}\Big(
\mbox{Tr}\Big(\frac{((JR)_0)^*}{r}(J_1(\varphi=2\pi)-J_1(\varphi=0))\Big)
+ 2 \pi r \Big\{ \frac{|c_2|}{r^2}
+\frac{|c_3|}{r^3}+...\Big\}\Big)=0,
\end{eqnarray*}
by Stokes's theorem, where $C_r$ denotes the circle enclosing the
disc $D_r$ of radius $r$, $\varphi$ is a coordinate on $C_r$, and
$|c_i|$ is the bound of $ \mbox{Tr}((JR)_i ^* \p_{\varphi}J) $,
$i=1,2,...\;$ . We have used the boundary condition \be
\lim_{r\rightarrow \infty}JR = (JR)_0 + (JR)_1(\varphi)r^{-1} +
O(r^{-2}), \ee which follows from (\ref{assympt}) for
$\hat{J}=JM$, and the fact that integrands are continuous on the
circle and hence bounded.  Since $ (JR)_0 $ is a constant matrix,
the first term in the series is a total derivative.

So far we have only used the assumption that $J$ is a solution of
(\ref{Wardeq}), but not that it has to be a uniton solution
defined by (\ref{n_uniton}).  Therefore, we have a more general
result for the total energy of a Ward solution of the form $\hat J
= JM$, where $J$ is an arbitrary solution to Ward equation.  Let
$\hat E$ and $E$ be the total energies of $\hat J$ and $J$
respectively, then
\be \label{energydiff} \hat E = E + \frac{(\mu
- \bar \mu)(1+|\mu|^2)}{|\mu|^2} \int_{\R ^2}
\mbox{Tr}(R[R_x,R_y]) \d x \d y. \ee
From this, the explicit
expression for the total energy of an n-uniton solution
(\ref{n_uniton}) follows.  First, consider a 1-uniton solution
$J_{(1)} = M_1$.  It can be written as $J_{(1)}= J_{(0)}M_1$,
where the constant matrix $J_{(0)}$,
 which satisfies (\ref{Wardeq}) trivially,  is chosen to
be the identity matrix.  Then, from (\ref{energydiff}), the total
energy of a 1-uniton solution is given by \be E_{(1)} = \frac{(\mu
- \bar \mu)(1+|\mu|^2)}{|\mu|^2} \int_{\R ^2} \mbox{Tr}(R_1[\p_x
R_1, \p_y R_1]) \d x \d y. \ee Therefore, using
(\ref{energydiff}),  we show by induction that the total energy of
an $n$-uniton solution (\ref{n_uniton}) is given by
\begin{eqnarray}
E_{(n)} &=& \frac{(\mu - \bar \mu)(1+|\mu|^2)}{|\mu|^2}
\sum^{n}_{k=1}\int_{\R ^2} \mbox{Tr}(R_k[\p_x R_k, \p_y R_k]) \d x
\d y \\ \nonumber &=& \pm 4 \pi \big( \frac{1+m^2}{m} \big) \sin (
\phi ) \; [\psi] \quad \left\{ \begin{array}{ll}
                       0 < \phi < \pi \\
                       \pi < \phi < 2 \pi,
                       \end{array}
                       \right.
\end{eqnarray}
where $\mu = m e^{i \phi} $, and we have used
(\ref{homotopy_of_psi}).  \koniec
We remark that the formula
(\ref{form_of_g}) reveals another topological interpretation of the
energy quantisation which is useful in practical calculations.
Consider the group element
(\ref{simple_elements}) with the index $k$ dropped. 
The Grassmanian projector $R$ 
in (\ref{n-uniton})
corresponds to a smooth map from the
compactified space to the projective space
$q:S^2\longrightarrow \CP^{N-1}$. The homotopy group
$\pi_2(\CP^{N-1})=\Z$ is non--trivial and the degree of $q$ is
obtained  by evaluating the homology class 
on a standard generator for $H^2(\CP^{N-1})$ represented
in a map  $q=(1, f_1, ..., f_{N-1})$
by the K\"ahler form 
\[
\Omega=-4i \partial\bar\partial \;{{\ln}}{(1+\sum_{j=1}^{N-1}|f_{j}|^2)}.
\]
This evaluation is just the integrating,
thus
\[
[q]=\frac{i}{8\pi}\int_{\R^2}q^*(\Omega).
\]
Evaluating the integrand we verify
that
\[
i\mbox{Tr}(R[R_x, R_y])=\frac{1}{4}q^*(\Omega).
\]
We conclude that the energy is proportional to the sum of the
topological degrees of Grassmanian projectors involved in the
definition of unitons.

In the remaining part of this Section we shall prove a Lorentz invariant generalisation of Theorem (\ref{main_th}).
We start off by looking at the quantisation of the momentum.
Following \cite{W88}, we have chosen the conserved energy functional for
a solution of (\ref{Wardeq}) to be that obtained from the
energy-momentum tensor of the associated standard chiral model.
However, for (\ref{Wardeq}) only the energy and the
$y$-component of momentum are conserved, while the $x$-component of
momentum is not.  The conserved $y$-momentum is given by
 \be
P=\int_{\R^2} {\cal P} \d x\d y , \ee
where the momentum density
is
\be \label{pdensity} {\cal P}= - \mbox{Tr}(J^{-1}J_tJ^{-1}J_y).
\ee
It turns out that this is also quantised and proportional to
the third homotopy class of the restricted extended solution.

\begin{prop} \label{momentum_prop}
The y-momentum of the $n$-uniton solution (\ref{n_uniton}) is given
by
\be \label{conj2} P_{(n)} = - 4 \pi \Big( \frac{1-m^2}{m}
\Big) |\sin (\phi)| \; [\psi].
\ee
Thus, unless $[\psi] = 0 $,  $ P=0 $ if and only if $m=1$.
\end{prop}

{\bf Proof.}
We first consider $\hat J=JM$ as in the the proof of Theorem \ref{main_th}.
The difference between the $y$-momentum densities (\ref{pdensity}) of
$ \hat J$ and $J$ is given by
\be \label{DeltaP}
\Delta \mathcal{P} \equiv \hat{\mathcal{P}} - \mathcal{P}
= \k \mbox{Tr} \Big( ({\bf 1}- \bar \k R )(J^{-1}J_tR_y + J^{-1}J_yR_t) +
\bar \k (R_yR_t) \Big).
\ee
Then the substitution
\begin{eqnarray*}
\mbox{Tr}(J^{-1}J_tR_yR)&=& \mbox{Tr} \Big((C-B) R_y \Big) \\
\mbox{Tr}(J^{-1}J_tRR_y)&=& \mbox{Tr} \Big( (B^*-C^*) R_y \Big)
\end{eqnarray*}
from the B{\"a}cklund
relations (\ref{backlund}) gives
\[
\Delta \mathcal{P} = \frac{\k}{\mu} \mbox{Tr} \Big( (1-|\mu|^2)
  R[R_x,R_y] + {\cal{T}} \Big),  \qquad \mbox{where}\qquad \k = \Big( 1 - \frac{\mu}{ \bar \mu } \Big).
\]
The term $ {\mathcal{T}} = \p_x (RJ^{-1}J_y) - \p_y (RJ^{-1}J_x)$
gives no contribution to the difference in the $y$-momenta
of $\hat J$ and $J$ as to the difference in the energies.  Thus we have a result for Ward solution of the form in $\hat J = JM$, where
$J$ is an arbitrary solution to Ward equation, its
$y$-momentum is given by
\be
\label{momdiff} \hat P = P - \frac{(\mu - \bar
\mu)(1-|\mu|^2)}{|\mu|^2} \int_{\R ^2} \mbox{Tr}(R[R_x,R_y]) \d x
\d y, \ee where $\hat P$ and $P$ are $y$-momenta of $\hat J$ and
$J$ respectively.

We then proceed by induction to obtain the expression for
the $y$-momentum of an $n$-uniton solution (\ref{n_uniton}), in the
same way as for the total energy. This gives
\begin{eqnarray}
P_{(n)} &=& -\frac{(\mu - \bar \mu)(1-|\mu|^2)}{|\mu|^2}
\sum^{n}_{k=1}\int_{\R ^2} \mbox{Tr}(R_k[\p_x R_k, \p_y R_k]) \d x
\d y \\ \nonumber &=& \mp 4 \pi \Big( \frac{1-m^2}{m} \Big) \sin
(\phi) \; [\psi] \quad \left\{ \begin{array}{ll}
                       0 < \phi < \pi \\
                       \pi < \phi < 2 \pi.
                       \end{array}
                       \right.
\end{eqnarray}
\koniec

We shall now exploit the $SO(1, 1)$ invariance of (\ref{Wardeq})
to combine Theorem \ref{main_th} and Proposition
\ref{momentum_prop} in the Lorentz invariant form
\begin{theo}
\label{theo_inv} For an $n$-uniton solution, the $SO(1, 1)$
invariant relation \be \label{invariant} E_{(n)}^2 - P_{(n)}^2 =
64\pi^2 \sin ^2 ( \phi) [\psi]^2 \ee holds.
\end{theo}
{\bf Proof.} Since the equation (\ref{Wardeq}) is invariant under
$SO(1,1)$, we can generate new solutions from a given one by
boosts in the $y-t$ plane.  In the coordinates $\{x, u=
\frac{1}{2}(t+y), v= \frac{1}{2}(t-y)\}$, the boosts are given by
$ x \rightarrow x, \quad u \rightarrow su, \quad v \rightarrow
s^{-1} v, \quad s \in \R^* $.  We shall show that a boost of an
$n$-uniton solution with a pole $\mu$ in the extended solution
gives rise to another $n$-uniton solution with the pole $\mu '= s
\mu$.

Consider the B{\"a}cklund relations (\ref{backlund}) expressed in
the $ \{x,u,v\}$ coordinates,
\begin{eqnarray} \label{bt}
( \mu R_x - R_u + RJ^{-1}J_u )({\bf 1}-R) &=& 0 \\
( \mu R_v - R_x + RJ^{-1}J_x )({\bf 1}-R) &=& 0. \nonumber
\end{eqnarray}
Let $J$ be an arbitrary solution of (\ref{Wardeq}), and $R(x,u,v)$
be the hermitian projector satisfying (\ref{bt}). Under the boost
to another solution $J \rightarrow J'$  we have $R \rightarrow R'
= R(x,su,s^{-1}v)$.  Changing the coordinates, we see that $R'$
will satisfy (\ref{bt}) with $\mu$ and $J$ replaced by $\mu '$
and $J'$, if $\mu ' = s \mu$.  That is, each restricted uniton
factor transforms as
\[
g_k = {\bf 1}+ \frac{\bar \mu-\mu}{\mu + \cot \Big(
\frac{\theta}{2} \Big)}R_k(x, u, v, \mu) \longrightarrow g_k'=
{\bf 1}+ \frac{ s \bar \mu-s\mu}{s\mu + \cot \Big(
\frac{\theta}{2} \Big)} R_k(x, su, s^{-1}v, \mu).
\]
Since boost is a continuous transformation it does not change the
homotopy types, and
\[
[\psi(x,u,v)] = [\psi(x,su,s^{-1}v)].
\]
Hence, under the transformation, $E_{(n)}$ and $P_{(n)}$ only
change due to the explicit factors of $\mu$ in (\ref{conj1}) and
(\ref{conj2}) respectively. The boosts rescale $\mu$ by $ m
\rightarrow sm$, keeping the phase $ \phi$ fixed.  This leads to
the $SO(1,1)$ invariance of $E_{(n)}^2 - P_{(n)}^2$. The formula
(\ref{invariant}) follows directly from (\ref{conj1}) and
(\ref{conj2}). \koniec

{\bf Examples.} Consider the $SU(2)$ case, where the third
homotopy class is equal to the topological degree and set $\mu=i$.
The uniton factors are of the form
\[
M_k= \frac{i}{1+|f_k|^2} \left (
\begin{array}{cc}
|f_k|^2-1&-2{f_k}\\
-2\ov{f_k}&1-|f_k|^2
\end{array}
\right ).
\]

{$\bf n=1.$}  In the one--uniton case $\p_tM_1=0$, and $M_1$ is
given by (\ref{M_k}) with $f_1=f_1({z})$ a rational function of
some fixed degree $Q$.  The energy density is
\[
{\cal E}_1=\frac{8|f_1'|^2}{(1+|f_1|^2)^2}=-i\mbox{Tr}\;(M_1[\p_z
M_1, \p_{\ov z}M_1])
\]
and $E=8\pi\;\mbox{deg}(g_1)$ in agreement with (\ref{conj1}). In
this case $g_1$ is a suspension of a rational map
$f_1:\CP^1\longrightarrow\CP^1$ and $\mbox{deg} \;(g_1)=
\mbox{deg} \;(f_1)$ is a simple illustration of the Freundenthal
Theorem which says that a suspension of maps of $d$--spheres
induces an isomorphism of  the homotopy groups.

{$\bf n=2.$} In the two--uniton case $M_1$ and $M_2$ are given by
(\ref{M_k}) with $\mu=i$ and
\[
q_1=(1, f),\qquad   q_2=(1+|f|^2)(1, f)-2i(tf'+h)(\ov{f}, -1),
\]
where $f$ and $h$ are rational functions of $z$ \cite{W95}. Define
$k=2(tf'+h)$.  The total energy density is \be {\cal
E}=\frac{8|(1+|f|^2)k'-2k\ov{f}f'|^2+16|kf'|^2+16(1+|f|^2)^2|f'|^2}{(|k|^2+(1+|f|^2)^2)^2}
\ee and
\[
E=\int_{\R_2} {\cal E} \;\d x\d y 
={8\pi}(\mbox{deg}(g_1)+\mbox{deg}(g_2))
\]
for all $t$. The quantisation of energy in this case has first
been observed in \cite{IM04}, where  it was shown that $E=8\pi Q$
where generically $Q=2\deg{f}+\deg{h}$. However,
$Q=\mbox{max}\;(2\deg{f}, \deg{h})$ if both $f$ and $h$ are
polynomials. Our formula (\ref{conj1})  is valid for all pairs
$(f, h)$.

\section{Conclusions}
\label{outlook} We have established the relation between the total
energy of time dependent solitons (\ref{n_uniton}) and homotopy
classes of associated extended solutions. To the best of our
knowledge this is the first example of a topological mechanism
ensuring the classical energy quantisation of moving solitons.

The $n$--uniton solutions (\ref{n_uniton}) form a subclass of all
finite energy solitons which satisfy the `trivial scattering'
boundary condition (\ref{trivial_scatt}). Dai and Terng
\cite{DT04} have demonstrated that the extended solution
corresponding to the general `trivial scattering' soliton has
poles at non--real points $\mu_1, ..., \mu_r$ with multiplicities
$n_1, ..., n_r$, and is a product of simple elements $G_{k, \a} \;
\a = 1, ..., r$ of the form in (\ref{n-uniton}). Our case
(\ref{n-uniton}) corresponds to $r=1$, but the method used in the
proof of Theorem \ref{main_th} applies to the general case as one
can choose a different $\mu$ at each iteration of the B\"acklund
transformations (\ref{backlund}). Formulae (\ref{form_of_g}) and
(\ref{energydiff}) lead to the general form of the total energy of
`trivial scattering' solitons
\begin{equation}
\label{more_general} E=4\pi\sum_{\alpha=1}^r\sum_{k=1}^{n_r}
\frac{1+m_\alpha^2}{m_\alpha}|\sin{\phi_{\alpha}}| [g_{k,
\alpha}],\qquad \mu_\alpha=m_\alpha e^{i\phi_\alpha},
\end{equation}
where
\[
g_{k,\alpha} = {\bf 1}+ \frac{\bar
\mu_\alpha-\mu_\alpha}{\mu_\alpha
 +\cot\Big(\frac{\theta}{2}\Big)}
R_{k,\alpha} \in U(N)
\]
and $R_{k,\alpha}$ are hermitian projections whose form is
determined by the B\"acklund relations. This agrees with the
result of Lechtenfeld and Popov \cite{LP01b}. The formula
(\ref{more_general}) is not directly linked to the homotopy type
of the extended solution, and the $SO(1, 1)$ invariance can not be
easily incorporated. This is why we have focused on the special
case (\ref{n_uniton}).

In $\cite{DM05}$ the $SU(2)$ integrable chiral model
(\ref{Wardeq}) has been analysed in the moduli space
approximation, when the time dependent slowly moving solitons
correspond to curves in the moduli space of static solitons which
are geodesic with respect to the natural metric
\[
h(\dot{\gamma}, \dot{\gamma})=\frac{1}{2}
\dot{\gamma}^p\dot{\gamma}^q\int_{\R^2}\frac{|\p_p f\ov{\p_q
f}|}{(1+|f|^2)^2} \d x\d y
\]
on the space of rational maps. Here $f=f(z, \gamma)$ is a rational
meromorphic function of $z=x+iy$ which depends on real parameters
(positions of zeroes and poles) $\gamma^p$, and $\p_p f=\p f/\p
\gamma^p$.

The kinetic energy of these approximate solitons is small, and
their total energy is  close (in the units of $8\pi$) to the
degree of the associated rational map. Theorem (\ref{main_th})
gives a class of exact solutions with quantised total energy, and
one may expect that the approximate  solitons of $\cite{DM05}$
arise from the time--dependent unitons by some limiting procedure.

\section*{Acknowledgements}
We  wish to thank Marcin Ka{\'z}mierczak, Nick Manton and Lionel
Mason for valuable comments, and Ivan Smith for clarifying some
aspects of homotopy theory.  We also thank the 
anonymous referee for valuable comments.
Prim Plansangkate is grateful to the
Royal Thai Government for funding her research.


\end{document}